# Relaxation of the Induced Orientational Order in the Isotropic Phase of Nematic Polymer

V.B. Rogozhin[a], S.G. Polushin[a]*, I.E. Lezova[a], G.E. Polushina[a], E.I. Ryumtsev[a], and N.A. Nikonorova[b].

[a]Faculty of Physics, St. Petersburg State University, St. Petersburg, Russia.

[b]Institute of Macromolecular Compounds of Russian Academy of Sciences, St. Petersburg, Russia.

*Email address for correspondence: s.polushin@spbu.ru*

**Abstract**

Orientational dynamics in the isotropic phase of a comb-shaped nematic polymer with mesogenic and functional side groups was studied using the Kerr effect and dielectric spectroscopy. For the first time, it was found that in a mesogenic polymer, in contrast to low-molecular-weight mesogens, the relaxation of the electric birefringence of a melt above the temperature of the nematic–isotropic phase transition can be presented by a sum of several exponential processes, two of which play a decisive role. These main processes replace each other in a temperature range of about 50 degrees. Dielectric spectroscopy also made it possible to distinguish two processes of orientational relaxation: the first is due to rotation of the side mesogenic groups, and the second is associated with motion of the main chain segments.

The isotropic phase of liquid crystals (LC) is characterized by a short-range order build-up as it approaches the temperature of phase transition ($T_c$) to LC state. The short-range order plays a significant role in a number of physical phenomena observed in the isotropic phase of LC, including static electric birefringence (EB or the Kerr effect) and EB relaxation in a pulsed electric field [1-6]. The discussed effects are typical for the temperature range of several tens of degrees higher than $T_c$, except for the region that is immediately close to phase transition, where order fluctuations become very big. These effects have been well studied, and the Landau – De Gennes theoretical model [7, 8] describing phase transitions in LCs has been successfully used in the analysis of pretransition phenomena in the isotropic phase of low molecular weight LC. According to the theory, EB relaxation time $\tau$ in an isotropic melt of LC depends on temperature as $\tau \propto \nu/(T-T^*)^\gamma$. Here, $\nu$ is the viscosity coefficient, $T^*$ is the temperature of the imaginary second-order phase transition, which is about 1 degree higher than $T_c$. In theory, the exponent $\gamma$ is equal to 1 that is confirmed by the experiment for low molecular weight LCs to within $10^{-2}$ [9]. LC polymers have a considerably more complex molecular architecture than low molecular weight LCs. The few studies of isotropic melts of comb-shaped LC polymers have shown that temperature dependence of static EB for polymers looks the same as for low molecular weight LCs: the Kerr constant *K* changes as $1/(T-T^*)^\gamma$, where γ is equal to 1 [10-12]. At the same time, it turned out that application of the Landau – De Gennes approach to describe EB dynamics in a polymer provides unexpected results. Thus, it was found that relaxation time calculated from rotational viscosity of a polymer in LC phase differs from τ measured in the experiment [11] by several times, and the exponent γ reaches 1.5 [13]. We have studied LC polymers with different molecular structures to determine the causes of such deviations [14]. It was found that γ is larger than 1 for all polymers, and with an increase of intra- and intermolecular interaction due to incorporation of functional (acid) side groups into a polymer structure, the exponent γ rises up to 3. It is important to note that other dynamic characteristics of polymer melts, which are not associated with the short-range order effects, such as viscosity, electrical conductivity and frequency dispersion of the Kerr constant are described in a standart way [14]. The results of analysis allow us to propose a model of orientational dynamics in LC polymers that explains the abnormally large value of the exponent γ by the replacement of one mechanism of rotational relaxation, where the polymer chains participate, by another mechanism, where the chains do not play a significant role, due to a temperature change of correlation length. The model implies a complicated nature of relaxation; therefore, the present work is focused on the analysis of EB relaxation spectrum to identify the contributions of various orientational mechanisms.

The electrooptical properties of the melt of comb-shaped nematic Cop-9.7 polyacrylate with cyanobiphenyl and carboxyl side groups were studied in this work [15,16]. Cyanobiphenyl mesogenic groups are responsible for the nematic phase formation. The strongly polar terminal –CN group provides a sufficient value of the electrooptical effect in the temperature range required for measurements. The copolymer contains 9.7 mol.% of functional acid –COOH groups, which form intra- and intermolecular hydrogen bonds and thereby affect the melt dynamics.

EB measurements were carried out in the Kerr cell with the electrodes of 0.4 cm long and the interelectrode gap of 0.03 cm. Pulsed electric fields with a strength up to 25 kV/cm and the electric pulse decay time of less than $10^{-7}$s were used.

The measured effect was characterized by the Kerr constant *K*, which is related to the induced birefringence Δ*n* and the electric field strength *E* by the Kerr law $K = \Delta n / E^2$.

Dynamics of the orientational macroscopic order, which appears in an isotropic-liquid medium as a response to an external field, is described by the equation $-\frac{\partial F}{\partial S} = \upsilon \frac{\partial S}{\partial t}$. Here, $F$ is the free energy, $t$ is the time, $S$ is the orientational order parameter. It means that after the field is turned off, the order relaxation occurs exponentially $S = S_0 \exp(-\frac{t}{\tau})$ with the characteristic time $\tau$. Change of the induced order in the experiment was recorded by optical anisotropy of the sample $\Delta n(t) = \Delta n_0 \exp(-\frac{t}{\tau})$. The relaxation time was determined as $\tau = \int_0^t \frac{\Delta n(t)}{\Delta n_0} dt$ from the obtained dependence of $\Delta n(t)/\Delta n_o$ on $t$ (inset in Fig. 1).

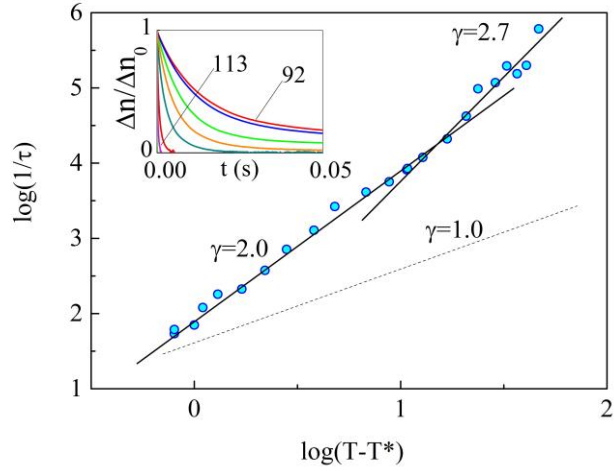

Fig. 1. Temperature dependence of the inverse integral relaxation time $\tau$ of the orientational order in the isotropic melt of Cop-9.7 polymer. The dashed line corresponds to the theoretical value $\gamma=1$. The inset shows the decay curves of the reduced birefringence in the temperature range from 113 to 92 °C.

The temperature dependence $1/\tau$ plotted on a log-log scale (Fig. 1) has the same abnormal behaviour as the previously studied LC polymers [13, 14]. The exponent $\gamma$ exceeds the theoretical value of 1 and is in the range of 2.0-2.7. The complex nature of relaxation process can be the reason for the abnormal relaxation. In this case, the photocurrent decay curve $f(t)$ can be described by a linear combination consisting of $i$ exponential terms: $f(t) = \sum_{i=1}^{N} a_i \exp(-\frac{t}{\tau_i})$, where $\tau_i$ are the partial relaxation times, $a_i$ are the partial amplitudes, $a_i=f_i(0)/\Sigma f_i(0)$. It is known that the exponential separation procedure is a non-trivial problem due to fundamental mathematical constraints, and in practice it is solved in various ways [17]. We used a common method, where experimental data is presented in the semi-log coordinates

$ln[\Delta n(t)/\Delta n_o]$ of *t*. With such plotting, a straight-line segment with the longest partial time $\tau_i$ is selected, then the data of this component is subtracted from the data array of the initial kinetic curve. Then the procedure is repeated many times to separate other exponents.

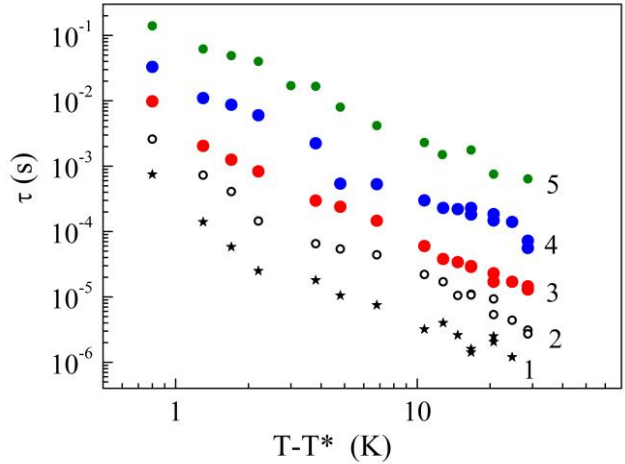

Fig. 2. Temperature dependence of the partial relaxation times $\tau_i$; i=1-5. The processes are marked with numbers. The main processes 3 and 4 are highlighted with large dots.

While separating the relaxation curve, one can obtain several exponential processes (Fig. 2) for which the partial amplitudes significantly differ (Fig. 3). Obviously, the key role is played by two processes with numbers 3 and 4, which differ from each other in time by almost an order of magnitude.

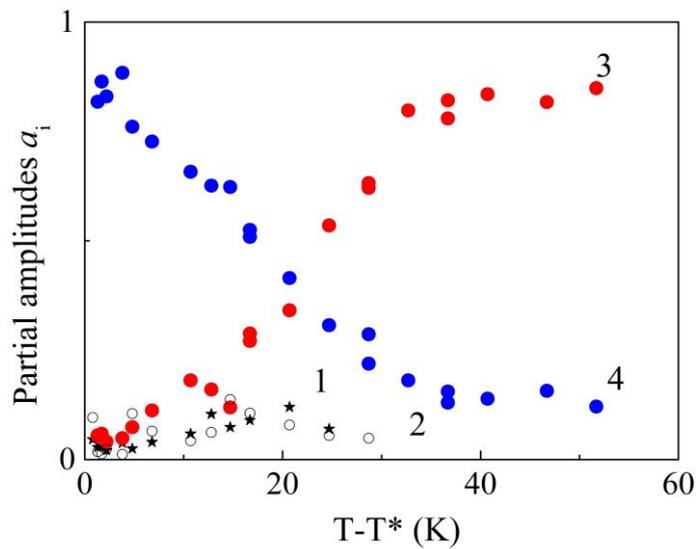

Fig. 3. The partial amplitudes $a_i$ in dependence to relative temperature for processes 1, 2, 3 and 4 in the isotropic melt of Cop-9.7.

The slower process 4 totally dominates in the vicinity of transition, at temperatures close to $T^* \approx T_c$. Its contribution decreases as the temperature rises. The fast process 3 becomes dominant at high temperatures. Such a regularity is well explained by the model proposed in [14]. Relaxation of the orientational order is related to disordering of nematic-type fluctuations when a field is turned off. The fluctuation sizes are given by the correlation length $r_c$ depending on temperature: $r_c = r_0[T^*/(T-T^*)]^{0.5}$. It is important that composition of the molecular ensemble involved in fluctuations inevitably changes with a change of $r_c$. Fluctuations reach a size of about 100Å at low temperatures [14]. This is noticeably larger than the distance between the neighboring chains, which is approximately equal to the length of the mesogenic groups - 20 Å, therefore, the chain segments are included in the ensemble together with the mesogenic groups. As a result, the relaxation time at low temperatures is determined by the slow motion of the polymer segments along with the mesogenic groups (process 4). The correlation length becomes less than 20Å at high temperatures (($T-T^*)\geq 20K$). At the same time, the fluctuation sizes are so small that only the orientational order carriers (the mesogenic groups) can be involved in them, which have a certain independence from the polymer chains due to the spacers $-(CH_2)_4-$. The motion of the mesogenic groups corresponds to a faster main process 3.

The dielectric relaxation times, both in the isotropic and in LC phase, were also measured. The dielectric spectra were obtained on Concept-21 Novocontrol Tecnologies dielectric spectrometer with Alpha-ANB high-resolution frequency analyzer in the frequency range $f$ from $10^{-1}$ to $2\times10^6$ Hz and the temperature range from 20 to 200$^0$C. The initial samples were the films obtained by pressing at a temperature several degrees higher than the temperature of isotropization $T_c$. Brass discs were used as the electrodes. The diameter of the upper disc was 20 mm. The sample thickness was set by 50-micron quartz fibers.

The dielectric spectra in the studied range of temperatures and frequencies — the frequency dependences of the dielectric loss factor $\varepsilon'' = \varphi(f)$ indicated the presence of two regions of maximum $\varepsilon''$, designated as α and δ processes, which are due to the relaxation processes of dipole polarization, that is confirmed by the maxima shift with increase of temperature towards high frequencies. The dielectric spectra were described by the empirical Havriliak-Negami equation (HN) [18]. The characteristic relaxation times $\tau_{max}$ were determined by the frequency position of $\varepsilon''_{max}$ according to the formula [19]:

$$\tau_{max} = \tau_{HN}\left[\frac{\sin(\frac{\pi(\alpha_{HN})\beta_{HN}}{2(\beta_{HN}+1)})}{\sin(\frac{\pi(\alpha_{HN})}{2(\beta_{HN}+1)})}\right]^{1/(\alpha_{HN})}.$$

The values of $\tau_{max}$ calculated by HN formula in the region of α and δ processes are given in Fig. 4.

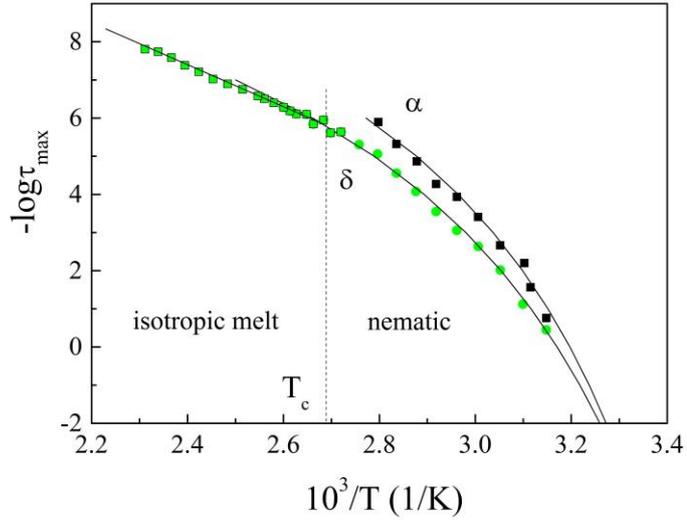

Fig. 4. Dependences of $-\log\tau_{max}$ on the inverse temperature for α и δ processes in the isotropic and LC phases of Cop-9.7.

Above $T_c$, the temperature dependence $\log\tau_{max}=\varphi(1/T)$ is linear and is described by the Arrhenius equation: $\tau(T)_{max} = \tau_0\exp(\frac{E_a}{RT})$, where $\tau_0=\tau_{max}$ at $T\to\infty$, $E_a$ is the activation energy, $R$ is the gas constant. The values of $-\log\tau_0$ and $E_a$ are 22±2 and 23±3 kcal/mol, respectively. Below $T_c$, the temperature dependences of $\log\tau_{max}$ for α and δ processes are curvilinear and are described by the empirical Vogel-Fulcher-Tammann-Hesse equation $\tau_{max} = \tau_0\exp(\frac{B}{T-T_0})$, where $\tau_0$, $B$ and $T_0$ are the temperature-independent parameters.

The non-linearity of $-\log\tau_{max}$ temperature dependences is typical for cooperative forms of molecular mobility, when the activation energy varies with temperature. The cooperative form of molecular mobility includes, particularly, the segmental mobility associated with transition to a high-elasticity state. Two cooperative processes, α and δ, were observed above the glass transition temperature for comb-shaped LC polymers, as well as for the polymer studied in this work, instead of one process (as in most polymers of other classes) [20-23].

The relaxation cooperativity along with the complex molecular architecture of comb-shaped LC polymers lead to difficulties in the interpretation of molecular mechanisms of dielectric relaxation. Usually, α-process is associated with the segmental mobility – the movement of individual kinetically independent chain sections (segments), consisting of several monomer units. The mesogenic group reorientation around the long axis also contributes to α-process. The second on a temperature scale is δ-process, which corresponds to the cooperative rotation of the mesogenic side groups around the short axis.

Cooperative relaxation processes represent the motion of relatively large parts of a molecule. The parameters of this motion are most sensitive to the chemical and morphological features of a polymer. It is clear from the above, why relaxation behaviour changes so much during the nematic–isotropic phase transition. In Fig. 4 one can see that the segmental motion (α-process) disappears in the isotropic phase, and the temperature dependence of δ-process becomes rectilinear. The motion of the side groups is strongly cooperated in LC phase in the presence of a long-range orientational order. In addition, there is a strong dynamic relation between the mesogenic fragments and the main chain. The mesogenic groups become more independent from each other and from the main chain with the disappearance of the long-range order.

Therefore, both dielectric spectroscopy and, for the first time, EB method, reveal the presence of two main relaxation processes in orientational dynamics of the comb-shaped LC polymer, which is in good agreement with our fluctuation model of the phenomenon [14]. One process is the motion of the mesogenic side groups around the short axis, while the other is due to the orientation of the main chain segments together with its environment. According to dielectric spectroscopy, the long-range nematic order contributes to the cooperative nature of the relaxation motion in LC phase. The strong influence of the short-range order on dynamics of the isotropic phase is evidenced by the Kerr effect.

References.
References.
1. J. Prost and J.R. Lalanne, Phys. Rev. **A 8**, 2090 (1973).

2. G.K.L. Wong and Y.R. Shen, Phys. Rev. **A l0**, 1277 (1974)

3. E.G. Hanson, Y.R. Shen, and G.K.L. Wong, Phys. Rev. **A 14**, 1281 (1976).

4. R. Yamamoto, S.Ishihara, and S.Hayakawa, Phys. Lett., A **69**, 276 (1978).

5. M.A. Agafonov, S.G. Polushin; T.A. Rotinyan, E.I. Rjumtsev, Kristallografiya, 31, 528 (1986).

6. E.I. Ryumtsev, S.G. Polushin, K.N. Tarasenko, and A.P. Kovshik, Zhurnal Fizicheskoi Khimii, **69**, 940 (1995).

7. P.G. de Gennes, Mol. Cryst. Liquid Cryst. **12**, 193 (1971).

8. P.G. de Gennes, Phys. Lett. **A 90**, 454 (1969).

9. J.J. Stankus, R. Torre, C.D. Marshall, S.R. Greenfield, A. Sengupta, A. Tokmakoff, M.D. Fayer, Chem. Phys. Lett. **194**, 213 (1992).

10. M. Eich, K. Ullrich, J.H. Wendorff, H. Ringsdorf, Polymer, **25**, 1271 (1984).

11. E.I. Rjumtsev, S.G. Polushin, K.N. Tarasenko, E.B. Barmatov, and V.P. Shibaev, Liquid Cryst., **21**, 777 (1996).

12. Th. Fuhrmann, M. Hosse, I. Lieker, J. Rubner, A. Stracke and J.H. Wendorff, Liquid Cryst., 26, 779 (1999).

13. V. Reys, Y. Dormoy, J. Gallani, P. Martinoty, P. Le Barny, J. Dubois, Phys. Rev. Lett., **61**, 2340 (1988).

14. S.G. Polushin; S.K. Filippov; T.S. Fiskevich; E.B. Barmatov, and E.I. Ryumtsev, Polymer Science, Ser. C, **52**, 24 (2010).

15. E.B. Barmatov, M.V. Barmatova, T.E. Grokhovskaya, V.P. Shibaev, Polymer Science A, **40**(7), 635 (1998).

16. S.G. Polushin, A.B. Melnikov, G.E. Polushina, E.B. Barmatov, V.P. Shibaev, A.V. Lezov, E.I. Rjumtsev, Polymer Science A, **43**, N5, 511 (2001).

17. A.A. Istratov, and O.F. Vyvenko, Revew of Scientific Instruments, **70**(2), 1233(1999).

18. S. Havriliak, S. Negami, Polymer, **8**, 161 (1967).

19. R. Diaz-Calleja, Macromolecules, **33**, 8924 (2000).

20. R. Zentel, G.R. Strobl, H. Ringsdorf, Macromolecules, **18**, 960 (1985).

21. G.S. Attard, G. Williams, G.W. Gray, D. Lacey, P.A. Gemmei, Polymer, **27**, 185 (1986).

22. K. Araki, Polymer J, **22**, 540 (1990).

23. N.A. Nikonorova, T.I. Borisova, V.P. Shibaev, Macromol.Chem.Phys., **201**, 226 (2000).